\documentclass[
aps,
prd,
showpacs,
preprint,
tightenlines,
nofootinbib,
floatfix]
{revtex4}
\usepackage{graphicx,
}
\begin{document}
%
\title{                 Solar neutrino oscillations\\
                and indications of matter effects in the Sun}
%
\author{        G.L.~Fogli}
\affiliation{   Dipartimento di Fisica
                and Sezione INFN di Bari\\
                Via Amendola 173, 70126 Bari, Italy\\}
\author{        E.~Lisi}
\affiliation{   Dipartimento di Fisica
                and Sezione INFN di Bari\\
                Via Amendola 173, 70126 Bari, Italy\\}
\author{        A.~Palazzo}
\affiliation{   Dipartimento di Fisica
                and Sezione INFN di Bari\\
                Via Amendola 173, 70126 Bari, Italy\\}
\author{        A.M.~Rotunno}
\affiliation{   Dipartimento di Fisica
                and Sezione INFN di Bari\\
                Via Amendola 173, 70126 Bari, Italy\\}

\begin{abstract}
Assuming the current best-fit solutions to the solar neutrino
problem at large mixing angle, we briefly illustrate how
prospective data from the Sudbury Neutrino Observatory (SNO) and
from the Kamioka Liquid scintillator Anti-Neutrino Detector
(KamLAND) can increase our confidence in the occurrence of
standard matter effects on active neutrino flavor oscillations in
the Sun, which are starting to emerge from current data.
\end{abstract}
\medskip
\maketitle

\section{Introduction}

The Chlorine \cite{Cl98}, Gallium \cite{Ab02,Ha99,Ki02},
Super-Kamiokande (SK) \cite{Fu01,Fu02} and Sudbury Neutrino
Observatory (SNO) \cite{AhCC,AhNC,AhDN} solar neutrino experiments
have convincingly established that the deficit of the observed
solar $\nu_e$ flux with respect to expectations \cite{Ba89,BP00}
implies new neutrino physics. In particular, the charged and
neutral current (CC and NC) data from SNO have proven the
occurrence of $\nu_e$ transitions into a different active state
$\nu_a$ with a statistical significance greater than $5\sigma$
\cite{AhNC}.

Barring sterile neutrinos and nonstandard $\nu$ interactions, such
transitions can be naturally explained by the hypothesis of flavor
oscillations \cite{Pont} in the $\nu_e\to\nu_a$ channel ($\nu_a$
being a linear combination of $\nu_\mu$ and $\nu_\tau$) driven by
nonzero $\nu$ squared mass difference and mixing angle parameters
($\delta m^2, \theta_{12}$) \cite{Maki}. The $(\nu_\mu,\nu_\tau)$
combination orthogonal to $\nu_a$ is probed by atmospheric $\nu$
oscillations \cite{Atmo}, with different parameters $(\Delta
m^2,\theta_{23})$ \cite{Revi}. The third mixing angle
$\theta_{13}$, needed to complete the $3\times 3$ mixing matrix,
is constrained to be small by additional reactor results
\cite{CHOO,Palo,Revi}, and can be set to zero to a good
approximation for our purposes.

The recent results from the Kamioka Liquid scintillator
AntiNeutrino Detector (KamLAND) \cite{KamL} have provided a
beautiful and crucial confirmation of the solar $\nu_e$
oscillation picture through a search for long-baseline
oscillations of reactor $\overline\nu_e$'s. The observed of
$\overline\nu_e$ disappearance in KamLAND has confirmed the
previously favored solution in the $(\delta m^2,\theta_{12})$
parameter space \cite{KamL}, often referred to as the large mixing
angle (LMA) region \cite{AhDN} in the literature (see, e.g.,
\cite{Last} and references therein). Moreover, the KamLAND data
have basically split this region into two allowed subregions,
which we will refer to as LMA-I and LMA-II, following
Ref.~\cite{KLou}. Although the LMA-I solution is favored by global
fits, the LMA-II solution at higher $\delta m^2$ is also
statistically acceptable, and we will discuss both cases in this work.%
\footnote{When the distinction between the LMA-I and LMA-II
solutions is irrelevant, we will refer to a generic ``LMA
solution.''}

Within the LMA region, solar neutrino oscillations are governed
not only by the kinematical mass-mixing parameters $(\delta m^2,
\theta_{12}$), but should also be significantly affected by the
interaction energy difference ($V=V_e-V_a$) between $\nu_e$'s and
$\nu_a$'s propagating in the solar (and possibly Earth) background
matter \cite{Matt,Barg}, through the so-called
Mikheev-Smirnov-Wolfenstein (MSW) mechanism \cite{Matt} in
adiabatic regime \cite{Adia}. Although Earth matter effects (i.e.,
day-night variations of solar event rates) remain elusive, solar
matter effects seem to emerge, at least indirectly, from the
combination of the available data (and especially from SNO),
through a preference for an average oscillation probability
smaller than $1/2$ at energies of a few MeV (see \cite{Smir} and
references therein).

The purpose of this article is to briefly illustrate how such
emerging indications of solar matter effects can be corroborated
in the LMA parameter region. In particular, we show that the
amplitude of matter effects (introduced as a free parameter
$a_\mathrm{MSW}$ in Sec.~II) can be significantly constrained by
using prospective data from SNO (Sec.~III) and KamLAND (Sec.~IV).
Both SNO and KamLAND can discriminate the case $a_\mathrm{MSW}=1$
(standard matter effects) against the case $a_\mathrm{MSW}=0$
(matter effects zeroed), and can thus
provide indirect indications for the MSW mechanism in the Sun.%
\footnote{A really ``direct'' evidence for MSW effects in the Sun
would require a full program of low-energy solar $\nu$
spectroscopy, probing the energy profile of the oscillation
probability down to the sub-MeV range \cite{Scho}.}

Although the occurrence of solar matter effects in the LMA region
is an unavoidable consequence of the standard model of electroweak
interactions, the importance of proving experimentally that ``they
are there'' and have the correct size cannot be overlooked.
Current and future research programs in neutrino physics,
including the accurate reconstruction of the kinematical mass and
mixing parameters for the three known generations of neutrinos,
and the associated searches for leptonic CP violation, largely
rely on our knowledge of the dynamical $\nu$ properties in matter.
Therefore, we think that increasing our confidence in the
occurrence of standard MSW effects in the solar interior is a
relevant (and reassuring) intermediate step towards the
realization
of these difficult and long-term research programs.%
\footnote{For similar reasons, an important goal of current and
future oscillation searches ``in vacuum'' is to observe a periodic
flavor change pattern, unavoidably associated to the mass-mixing
parameters.}

\section{Switching matter effects on and off}

In the active two-flavor oscillation picture, the solar $\nu$
evolution equation in the space coordinate $x$ reads
\begin{equation}\label{H}
  i\frac{d}{dx}\left(\begin{array}{c}\nu_e\\ \nu_a\end{array}\right)=
  {\cal H}\left(\begin{array}{c}\nu_e\\ \nu_a\end{array}\right)
\end{equation}
where the Hamiltonian $\cal H$ can be split into kinematical
\cite{Pont,Maki} and dynamical \cite{Matt} components, describing
the free and interaction $\nu$ energy, respectively,
\begin{equation}\label{Hsplit1}
  {\cal H}={\cal H}_\mathrm{kin}+{\cal H}_\mathrm{dyn}\ ,
\end{equation}
with
\begin{equation}
\label{Hkin}
  {\cal H}_\mathrm{kin} = \frac{k}{2}
  \left(\begin{array}{cc}-\cos2\theta_{12}&\sin2\theta_{12}\\
  \sin2\theta_{12}&\cos2\theta_{12}\end{array}\right)\ ,
\end{equation}
and
\begin{equation}
  \label{Hdyn}
  {\cal H}_\mathrm{dyn}=\frac{V(x)}{2}
  \left(\begin{array}{cc} 1 & 0\\
  0&-1\end{array}\right)\ .
\end{equation}
In the above equations,  $E$ is the  neutrino energy, $k=\delta
m^2/2E$ is the neutrino oscillation wavenumber, and $V(x)$ is the
difference between the $\nu_e$ and $\nu_a$ interaction energies
with ordinary matter \cite{Matt} at the position $x$,
characterized by the electron number density $N_e$,
\begin{equation}\label{V}
  V(x)=\sqrt{2} G_F N_e(x)\ .
\end{equation}

The relevance of the dynamical term ${\cal H}_\mathrm{dyn}$ on the
$\nu_e$ survival probability $P_{ee}$  strongly depends on the
oscillation parameter values. As it is well known, for $\delta
m^2\lesssim 10^{-9}$ eV$^2$ dynamical effects are small or
negligible (so-called quasivacuum and vacuum oscillation regimes),
while for $\delta m^2\gtrsim 10^{-7}$ eV$^2$ they are definitely
relevant (so-called MSW regime) \cite{Revi}. The available solar
neutrino data favor solutions in the MSW regime (and in particular
the LMA region of parameters \cite{AhDN}), but do not exclude
(quasi)vacuum solutions with sufficiently high confidence
\cite{Last,QVO1,QVO2}. In other terms,  cases where one can
phenomenologically set ${\cal H}_\mathrm{dyn}\simeq 0$ are not
ruled out by current solar neutrino data alone, implying that no
compelling evidence for matter effects has been found so far in
this data set.

The first KamLAND data \cite{KamL}, however, have excluded
(quasi)vacuum solutions to the solar neutrino problem, and have
unambiguously selected the LMA region of the parameter space
\cite{KLou}. Therefore, we can restrict ourselves to relatively
high values of $\delta m^2$ (say, above $10^{-5}$ eV$^2$), where
matter effects are expected to play a relevant role. It makes then
sense to ask whether the data, by themselves, can globally provide
some evidence that matter effects are really there (${\cal
H}_\mathrm{dyn}\neq 0$) and have their expected size
[Eq.~(\ref{V})]. One can rephrase this question by introducing a
free parameter $a_\mathrm{MSW}$ modulating the overall amplitude
of the interaction energy difference $V$ in the dynamical term
${\cal H}_\mathrm{dyn}$,
\begin{equation}
\label{Ha} V \to a_\mathrm{MSW}\cdot V\ ,
\end{equation}
so that standard matter effects can be formally switched on and
off by setting $a_\mathrm{MSW}=1$ and $a_\mathrm{MSW}=0$,
respectively. One can then try to check whether the data prefer
the first or the second option for $a_\mathrm{MSW}$. Furthermore,
by treating $a_\mathrm{MSW}$ as a continuous parameter, one can
try to constrain its allowed range through global data analyses: A
preference for $a_\mathrm{MSW}\sim O(1)$ would then provide an
indirect indication for the occurrence of matter effects with
standard size, as opposed to the case of pure ``vacuum''
oscillations (${\cal H}_\mathrm{dyn}\simeq 0$).%
\footnote{A similar approach has been used in the context of
atmospheric $\nu_\mu\to\nu_\tau$ oscillations, in order to find
indirect indications for the expected $L/E$  oscillation pattern
\cite{Nexp}. In that case, a continuous free parameter $n$ has
been formally introduced as an energy exponent ($L\cdot E^n$), and
a strong preference of the data for $n\simeq -1$ has been found
\cite{Nexp}, supporting the standard $\nu_\mu\to\nu_\tau$
oscillation picture.}

We have verified that the current solar neutrino data, by
themselves, place only very loose and uninteresting  limits on
$a_\mathrm{MSW}$, as far as the mass-mixing oscillation parameters
are left unconstrained. In fact, since the oscillation physics
depends mostly on the ratio $V/k$, a variation of the kind $V\to
a_\mathrm{MSW}V$ is largely absorbed by a similar rescaling of $k$
(i.e., of $\delta m^2$). In order to break this degeneracy, we
need to include explicitly an experiment which is highly sensitive
to $\delta m^2$ and basically
insensitive to matter effects, such as KamLAND.%
\footnote{For the sake of consistency, we will include Earth
matter effects with variable $a_\mathrm{MSW}$ also in the analysis
of current or prospective KamLAND data (Sec.~III). Such effects
become formally nonnegligible only for large values of
$a_\mathrm{MSW}$.}
We will thus focus, in the following, on the $\delta m^2$ values
selected by reactor neutrino data in the LMA range (roughly
$10^{-4\pm 1}$ eV$^2$), and eventually on the specific LMA-I and
LMA-II solutions favored by current global fits including KamLAND
\cite{KLou}.

\section{Matter effects and the SNO CC/NC double ratio}

Let us restrict the analysis to the LMA region, whose best-fit to
solar neutrino data alone, as taken from \cite{Last}, is reached
at $\delta m^2=5.5\times 10^{-5}$ eV$^2$ and
$\sin^2\theta_{12}=0.3$. Within this region, current solar
neutrino data from SK and SNO provide already some indirect
indications in favor of matter effects in the Sun, through their
preference for $P_{ee}\sim 1/3< 1/2$, where $P_{ee}$ is the
average $\nu_e$ survival probability in the SK-SNO energy range
\cite{Gett,Bere,Impr}.

Indeed, in the LMA region and for $a_\mathrm{MSW}=1$ (standard
matter effects), adiabatic MSW transitions \cite{Matt,Adia} occur
in the Sun, leading to a survival probability of the form (up to
residual Earth matter effects):
\begin{equation}\label{Pa1}
P_{ee}\simeq \cos^2\theta_{12}^\prime \cos^2\theta_{12} +
\sin^2\theta_{12}^\prime \sin^2\theta_{12}\ \ (a_\mathrm{MSW}=1)\
,
\end{equation}
where $\theta_{12}^\prime$ is the rotation angle which
diagonalizes $\cal H$ at the $\nu_e$ production point in the solar
core. On the other hand,  for hypothetically zeroed matter effects
($a_\mathrm{MSW}=0$), one would get an energy-independent form for
$P_{ee}$ in the LMA region,
\begin{equation}\label{Pa0}
P_{ee}\simeq 1-\frac{1}{2}\sin^2 2\theta_{12}\ \
(a_\mathrm{MSW}=0)\ ,
\end{equation}
as originally suggested by Gribov and Pontecorvo \cite{Grib} prior
to the MSW papers \cite{Matt} (see also \cite{Gosw,Pena}).

In the SNO energy range $(E\gtrsim 5$~MeV), the above two
expressions lead to comparable results in the second octant of the
mixing angle $(\theta_{12}>\pi/4)$, but differ considerably in the
first octant, where $P_{ee}(a_\mathrm{MSW}=1)< 1/2$, while
$P_{ee}(a_\mathrm{MSW}=0)>1/2$. Since the LMA likelihood extends
only marginally in the second octant \cite{Last}, there are very
good chances that SNO can discriminate the cases
$a_\mathrm{MSW}=0$ and $a_\mathrm{MSW}=1$ through the double ratio
of experimental-to-theoretical CC and NC events, which is
SSM-independent, and is equivalent to the average of $P_{ee}$ over
the SNO energy response function \cite{Gett,Smir,Vill,PeMa}.

Figure~1 shows isolines of the CC/NC double ratio in the usual
mass-mixing plane,%
\footnote{In Fig.~1, the choice of a linear scale for
$\sin^2\theta_{12}$ enhances the large mixing region, at the cost
of ``squeezing'' the (currently excluded) region of small mixing
angles. Among the three $\delta m^2$ decades shown, the middle one
is relevant for the LMA solution.}
for both $a_\mathrm{MSW}=1$ and $a_\mathrm{MSW}=0$, using the
current SNO CC threshold \cite{AhNC}.  It is evident from this
figure that, by excluding CC/NC values greater than 1/2 with high
confidence, the SNO experiment can conclusively discriminate the
cases of standard and zeroed matter effects, and will provide two
very useful (correlated) indications, namely: (1) that
$\theta_{12}<\pi/4$; and (2) that matter effects indeed take place
in the Sun. To reach this conclusion one needs only to know, in
addition, that the oscillation parameters are roughly in the LMA
region---a piece of information which has been indeed provided by
the first KamLAND data.

Although such simple considerations arise from well-known
properties of the oscillation probability \cite{Gosw,Pena}, we
think that the crucial role of {\em future\/} SNO CC/NC data
\cite{Hall} in establishing the occurrence of matter effects in
the Sun has perhaps not been stressed enough. Let us review, in
fact, the current situation. Within the LMA region, neither SK nor
the Gallium experiments can really discriminate the two octants of
$\theta_{12}$ at present (see, e.g., Fig.~2 in \cite{Gett}), and
cannot individually prove that solar matter effects are taking
place. The Chlorine experiment \cite{Cl98}, which observes an
event rate suppression of $\sim 1/3$ as compared to standard solar
model (SSM) predictions \cite{BP00}, prefers the first octant and
thus the presence of matter effects, as it is well known; however,
this indication is unavoidably SSM-dependent and thus not totally
compelling, especially if additional (hypothetical) experimental
systematics are invoked \cite{Gosw,Pena}. A SSM-independent
preference for $P_{ee}<1/2$ has been provided first by the
combination of SNO CC and SK data \cite{AhCC} and then by SNO data
alone through the CC/NC double ratio \cite{AhNC}, but not yet with
a significance high enough to rule out $P_{ee}=1/2$ \cite{Gett}.
Let us consider, in particular, the latest SNO constraints in the
plane $(\Phi_e,\Phi_{\mu\tau})$ charted by the solar $\nu_e$ and
$\nu_{\mu,\tau}$ fluxes, as shown in Fig.~3 of the original SNO
paper \cite{AhNC}. In such a figure, although the SNO best-fit
point clearly prefers $P_{ee}\sim 1/3$ (corresponding to
 $\Phi_{\mu\tau}\simeq 2\Phi_e$), the 95\% C.L.\
ellipse is still compatible with $P_{ee}\sim 1/2 $ (namely,
$\Phi_{\mu\tau}\simeq \Phi_e$). However, future SNO NC and CC data
 can considerably improve the constraints on $P_{ee}$, by
reducing both the statistical and the systematic error on the
CC/NC ratio \cite{Hall}. In particular, the current
anticorrelation between the CC and NC event rate uncertainties,
which prevents a significant cancellation of errors in the CC/NC
ratio, will be largely suppressed by the future event-by-event
reconstruction of the NC data sample \cite{Hall}.

In conclusion, although the combination of all current solar
neutrino data suggests a pattern of $P_{ee}$ compatible with the
LMA energy profile \cite{Bere,Impr} and indicates an overall
preference for the first octant of $\theta_{12}$ \cite{AhDN}, the
emerging indications in favor of solar matter effects from this
data set are not strongly compelling yet. Among the solar neutrino
experiments, in the near future only SNO appears to be able to
improve significantly this situation through new CC and NC data
\cite{Hall}, which can discriminate $a_\mathrm{MSW}=1$ from
$a_\mathrm{MSW}=0$ by excluding $P_{ee}$ values greater than $1/2$
in the LMA region, as we have tried to emphasize in this
Section.%
\footnote{In the presence of $3\nu$ mixing ($\theta_{13}\neq 0$),
this requirement would become slightly more stringent: SNO should
prove that $P_{ee}<s^4_{13}+c^4_{13}/2\leq 0.453$, where the
global $3\sigma$ upper limit on $s^2_{13}=\sin^2\theta_{13}$
($s^2_{13}\leq 0.05$ \cite{Last}) is assumed.}
At the same time, upper bounds on CC/NC smaller than $1/2$ will be
helpful to strengthen the upper limits on $\delta m^2$, as evident
from the left panel in Fig.~1 (see also \cite{PeMa}). Should
instead future SNO data drive the preferred value of $P_{ee}$ from
$\sim 1/3$ to relatively higher values, it would clearly become
much more difficult to assess the occurrence of MSW effects in the
Sun.

\section{MATTER EFFECTS IN GLOBAL ANALYSES INCLUDING KAMLAND}

In the previous Section, we have briefly illustrated how a single
datum (the SNO CC/NC double ratio) can discriminate the case of
standard matter effects $(a_\mathrm{MSW}=1)$ from the case of
zeroed matter effects $(a_\mathrm{MSW}=0)$ in the LMA parameter
region. By using further experimental information from KamLAND,
one could try to test whether the ``solar~+~KamLAND'' combination
of data can constrain matter effects in the Sun to have the right
size [$a_\mathrm{MSW}\sim O(1)$]. In this kind of analyses,
KamLAND basically fixes the oscillation parameters $(\delta
m^2,\sin^2\theta_{12})$, and thus the kinetic part of the
Hamiltonian, ${\cal H}_\mathrm{kin}$ [Eq.~(\ref{Hkin})]. The role
of solar neutrino data is then to check that the overall amplitude
$a_\mathrm{MSW}$ of the interaction energy difference $V$ in the
dynamical term ${\cal H}_\mathrm{dyn}$
[Eqs.~(\ref{Hdyn}--\ref{Ha})] is consistent with the standard
electroweak model ($a_\mathrm{MSW}=1$).

We have thus performed global analyses including both current
solar neutrino data and current (or prospective) KamLAND data,
with $(\delta m^2,\sin^2\theta_{12},a_\mathrm{MSW})$
unconstrained.%
\footnote{We refer the reader to \cite{Gett} and \cite{Last,KLou}
for technical details of our solar and KamLAND data analysis,
respectively.}
In particular, the analysis of current KamLAND data is based on
the binned energy spectrum of reactor neutrino events observed
above 2.6 MeV (54 events) \cite{KamL}. Prospective KamLAND
spectral data have instead been generated, with the same energy
threshold and binning, by assuming either the LMA-I best-fit point
($\delta m^2=7.3\times 10^{-5}$ eV$^2$ and
$\sin^2\theta_{12}=0.315$) or the LMA-II best-fit point ($\delta
m^2=15.4\times 10^{-5}$ eV$^2$ and $\sin^2\theta_{12}=0.300$)
\cite{KLou}, and increased statistics ($5\times 54$ and $10\times
54$ events). The CHOOZ reactor data \cite{CHOO} are also included.

Figure~2 and 3 show the results of such global fits, in terms of
the function $\Delta \chi^2=\chi^2-\chi^2_{\min}$  for variable
$a_\mathrm{MSW}$ and unconstrained (i.e., minimized away)
mass-mixing parameters. The $n\sigma$ bounds on $a_\mathrm{MSW}$
are then given by $\Delta \chi^2=n^2$. Let us focus first on the
solid curve, which  refers to the fit with {\em current\/} KamLAND
data, and is identical in both Figs.~2 and 3. It appears that such
curve can already place $>3\sigma$ upper and lower bounds on
$a_\mathrm{MSW}$.  In particular, the hypothetical case of zeroed
matter effects is already disfavored at $\sim 3.5\sigma$, thus
providing an indirect indication in favor of matter effects in the
Sun. The best-fit value of $a_\mathrm{MSW}$ is close to the
standard prediction ($a_\mathrm{MSW}=1$). However, there are also
other quasi-degenerate minima, and the overall $\pm 3\sigma$ range
for $a_\mathrm{MSW}$, spanning about three orders of magnitude, is
rather large. The width of this range can be understood by
recalling the following facts: (1) the LMA range of $\delta m^2$
constrained by solar neutrino data, which covers about one decade
\cite{Gett,Last}, can be shifted up or down by shifting
$a_\mathrm{MSW}$ with respect to 1, since the LMA oscillation
physics depends on $V/k\propto a_\mathrm{MSW}/\delta m^2$; (2) the
range of $\delta m^2$ constrained by current terrestrial data
(including KamLAND+CHOOZ) data, which covers about two decades
\cite{KLou}, is much less affected by $a_\mathrm{MSW}$ variations.
As a consequence, by appropriately shifting $a_\mathrm{MSW}$, it
is possible to overlap the reconstructed ranges of $\delta m^2$
from solar and from reactor data over about $1+2$ decades. When
the overlap sweeps through the degenerate $\delta m^2$ intervals
allowed by KamLAND alone \cite{KLou}, the fit is locally improved,
leading to a ``wavy'' structure in the $\Delta\chi^2$. In
conclusion, although current solar+reactor data strongly disfavor
$a_\mathrm{MSW}=0$ (zeroed matter effects) and provide a best fit
close to $a_\mathrm{MSW}=1$ (standard matter effects), the
presence of other local minima in the $\Delta\chi^2$ function, as
well as the broad $3\sigma$ allowed range for $a_\mathrm{MSW}$, do
not allow to claim a clear evidence of standard matter effects
from current data.

The previous qualitative arguments about the overlap of the
reconstructed ranges of $\delta m^2$ from solar and reactor data
fits (with variable $a_\mathrm{MSW}$) also provide a clue to what
one should expect with future KamLAND data. With increasing
statistics, the KamLAND reconstructed range of $\delta m^2$ will
shrink from two decades to a fractionally small value,%
\footnote{Provided that $\delta m^2\lesssim 2\times 10^{-4}$
eV$^2$ (see, e.g., \cite{Last}). This condition is fulfilled by
the LMA-I best-fit point and, to a large extent, also by the
LMA-II point.}
so that the overlap with the $\delta m^2$ range constrained by
current solar neutrino data will also be reduced from the current
``$1+2$'' decades to prospective ``1+0'' decades. Therefore, for
both the LMA-I and the LMA-II case, we expect that
$a_\mathrm{MSW}$ can be constrained within about one order of
magnitude by future KamLAND data. These expectations are
quantitatively verified by the broken curves in Figs.~2 and 3, as
we now discuss.

The broken curves in Fig.~2 refer to prospective KamLAND data,
generated by assuming as true solution the LMA-I best-fit point.
The energy threshold, the binning, and the systematic
uncertainties are assumed to be the same as for the current
KamLAND data. The dotted (dashed) curve refers to a number of
reactor neutrino events five (ten) times larger than the current
statistics, $N=54$. It can be seen that the global fit will
progressively constrain $a_\mathrm{MSW}$ within one decade at $\pm
3\sigma$ and, most importantly, will lead to a marked preference
for $a_\mathrm{MSW}\simeq 1$, which is not yet evident in the
present data (solid curve). A further increase of the simulated
KamLAND data sample will not lead to significantly more stringent
bounds on $a_\mathrm{MSW}$, the fit being then dominated by the
solar neutrino data uncertainties. In conclusion, if the LMA-I
solution is the true one, there are good prospect to test
unambiguously the occurrence and size of standard matter effects
in the Sun with higher KamLAND statistics. The uncertainty on
$a_\mathrm{MSW}$ will eventually be saturated by the solar $\nu$
uncertainties.%
\footnote{As previously remarked, new solar $\nu$ data, especially
from SNO, might reduce such uncertainties. However, it seems to us
premature to study in detail also the effect of prospective solar
neutrino data from current and future experiments.}

Figure~3 is analogous to Fig.~2, but the broken curves refer now
to KamLAND data simulated for the LMA-II solution. Two important
differences emerge by a comparison of the prospective fits to
$a_\mathrm{MSW}$ in Figs.~2 and 3, namely, a shift of the best-fit
value, and a relaxation of the upper bounds for the LMA-II case.
The shift is a reflection of the ``mismatch'' between the LMA-II
value $\delta m^2=15.4\times 10^{-5}$ eV$^2$ \cite{KLou} and the
value $\delta m^2=5.5\times 10^{-5}$ eV$^2$ preferred by current
solar neutrino data \cite{Last}, which is traded for an increase
of $a_\mathrm{MSW}$, when this parameter is left free. The
relaxation of the $a_\mathrm{MSW}$ upper bound in Fig.~3 is
instead due to the vicinity of the LMA-II  $\delta m^2$ value to
the critical onset of average oscillations in KamLAND ($\sim
2\times 10^{-4}$ eV$^2$). However, despite these drawbacks, the
prospective allowed range of $a_\mathrm{MSW}$ for the LMA-II case
in Fig.~3 will be still compatible with standard matter effects at
the $\sim 2\sigma$ level.

From Fig.~3 we also learn a more general lesson. Any mismatch
between the oscillation parameters (especially $\delta m^2$) as
separately reconstructed by KamLAND data and by solar data will
lead to deviations from the standard oscillation picture, if
allowance is given the fitting model. In our case, the deviation
will show up as a slightly nonstandard matter effect
($a_\mathrm{MSW}\neq 1$) in the LMA-II case. If such a situation
occurs in the future, it might be difficult to assess whether this
kind of ``deviation'' is due to statistical fluctuations, or
rather to new neutrino interactions (e.g., nonuniversal neutral
current couplings in the $\nu_e-\nu_a$ sector). Therefore,
selecting between the LMA-I and LMA-II solution is of crucial
importance to test any subleading effects on solar neutrinos,
beyond standard adiabatic oscillations in matter. While a
confirmation of the LMA-I case will probably lead to tight upper
bounds on any subleading effect, in the LMA-II case one should
expect such effects to be slightly favored by global fits.

\section{Summary and conclusions}

In the simplest picture, solar neutrino oscillations depend on the
kinematical parameters $(\delta m^2,\sin^2\theta_{12})$ and on
standard dynamical MSW effects in matter. The knowledge of the
kinematical mass-mixing parameters has been enormously improved
after the first KamLAND data, which have restricted their range
within the so-called LMA region and, in particular, in two
subregions named LMA-I and LMA-II. Standard dynamical effects in
current solar neutrino data are starting to emerge through an
increasingly marked preference for $P_{ee}<1/2$, but still remain
not clearly identified and partly elusive.

In order to quantify statistically the occurrence of MSW effects,
we have introduced a free parameter $a_\mathrm{MSW}$ modulating
the amplitude of the $\nu$ interaction energy difference in the
neutrino evolution equation, the cases $a_\mathrm{MSW}=1$ and
$a_\mathrm{MSW}=0$ corresponding to the standard and
(hypothetically) zeroed MSW effect, respectively. The SNO double
ratio of CC/NC events can clearly discriminate, in a
SSM-independent way, the case $a_\mathrm{MSW}=1$ against
$a_\mathrm{MSW}=0$, provided that the current indication in favor
of $P_{ee}<1/2$ is confirmed with higher statistical significance.

By treating $a_\mathrm{MSW}$ as a continuous parameter, we have
then performed a global analysis including current solar, CHOOZ,
and KamLAND data. The results are encouraging, since upper and
lower bounds on $a_\mathrm{MSW}$ appear to emerge at the
$>3\sigma$ level. In particular, the case of ``zeroed'' matter
effects is significantly disfavored. Moreover, the best-fit is
tantalizingly close to the standard expectations for matter
effects ($a_\mathrm{MSW}=1$). However, the presence of other
quasi-degenerate minima, and the very wide allowed range for
$a_\mathrm{MSW}$ (spanning about three decades at the $3\sigma$
level) prevent any firm conclusion about the occurrence of
standard matter effects at present.

The situation will greatly improve, even with unaltered solar
neutrino data, through higher KamLAND statistics (say, by a factor
of five or ten, as considered in Figs.~2 and 3). In both  the
LMA-I and LMA-II cases, it appears possible to reduce the current
uncertainty on $a_\mathrm{MSW}$ by about two orders of magnitude.
The prospects are particularly promising for the LMA-I solution.
In the LMA-II case, in fact, the reconstructed parameter
$a_\mathrm{MSW}$ might be biased towards higher values than the
standard expectation, as a result of a slight mismatch between the
solar and KamLAND reconstructed value of $\delta m^2$. Therefore,
the selection of a single solution in the LMA oscillation
parameter space appears to be crucial, before any definite
conclusion can be made on the emerging indications of standard
matter effects in the Sun.

\acknowledgments This work is supported in part by
the Istituto Nazionale di Fisica Nucleare (INFN) and by the
Italian Ministry of Education (MIUR) through the ``Astroparticle
Physics'' project. We thank A.\ Marrone and D.\ Montanino for
useful discussions and suggestions.



\begin{figure}
\vspace*{-0cm}\hspace*{-2.2cm}
\includegraphics[scale=0.9, bb= 30 100 500 700]{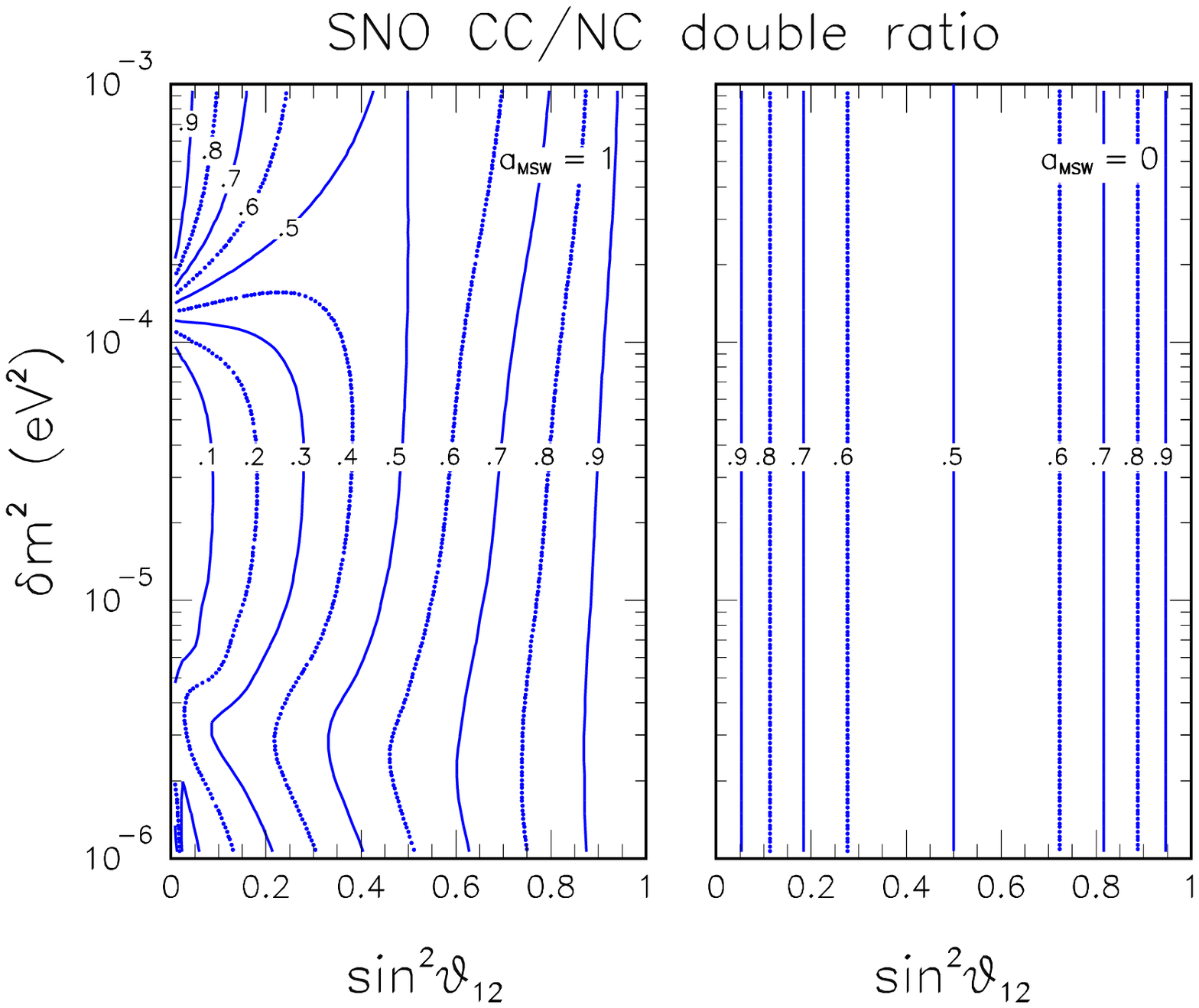}
\vspace*{-1cm} \caption{\label{fig1} The SNO CC/NC double ratio
for standard and zeroed matter effects ($a_\mathrm{MSW}=1$ and 0,
respectively). The parameter $a_\mathrm{MSW}$ is conventionally
introduced to modulate the standard amplitude of the $\nu$
interaction energy difference $V=\sqrt{2}\,G_F\, N_e$ in matter
($V\to a_{\mathrm{MSW}}\, V$). CC/NC values lower than 0.5, being
reachable  for $a_\mathrm{MSW}=1$ (but not for $a_\mathrm{MSW}=0$)
are clearly indicative of the occurrence of matter effects in the
LMA region. The exclusion of CC/NC values greater than 0.5 with
high statistical significance is thus an important future goal for
SNO.}
\end{figure}

\begin{figure}
\vspace*{-0cm}\hspace*{-2.2cm}
\includegraphics[scale=0.9, bb= 30 100 500 700]{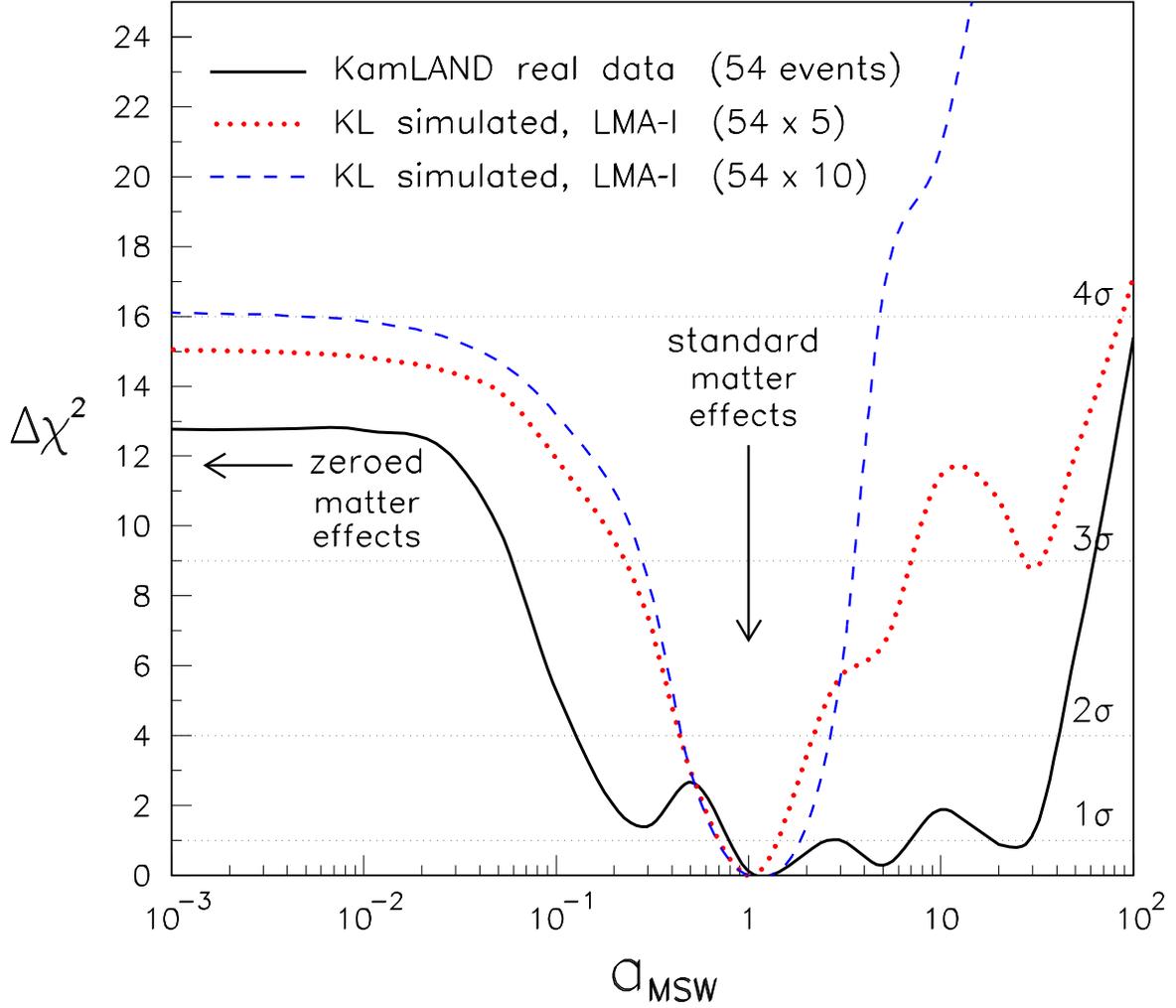}
\vspace*{-1cm} \caption{\label{fig2} Bounds on $a_\mathrm{MSW}$
(considered as a continuous free parameter) for unconstrained
$(\delta m^2,\,\sin^2\theta_{12})$,  including current solar and
CHOOZ neutrino data, as well as current or prospective KamLAND
data. The solid curve refers to the fit including {\em current\/}
KamLAND spectrum data above 2.6 MeV threshold (54 events
\cite{KamL}), and shows that the hypothetical case of zeroed
matter effects is already significantly disfavored. The other
curves refer {\em simulated\/} KamLAND data, generated by assuming
the LMA-I solution of Ref.~\cite{KLou}, and statistics increased
by a factor of five (dotted curve) and of ten (dashed curve). The
marked preference for $a_\mathrm{MSW}\simeq 1$ illustrates the
possibility of assessing the standard size of solar matter effects
within a factor of $\sim 2$ in future global analyses, provided
that LMA-I solution is correct. See the text for details.}
\end{figure}

\begin{figure}
\vspace*{-0cm}\hspace*{-2.2cm}
\includegraphics[scale=0.9, bb= 30 100 500 700]{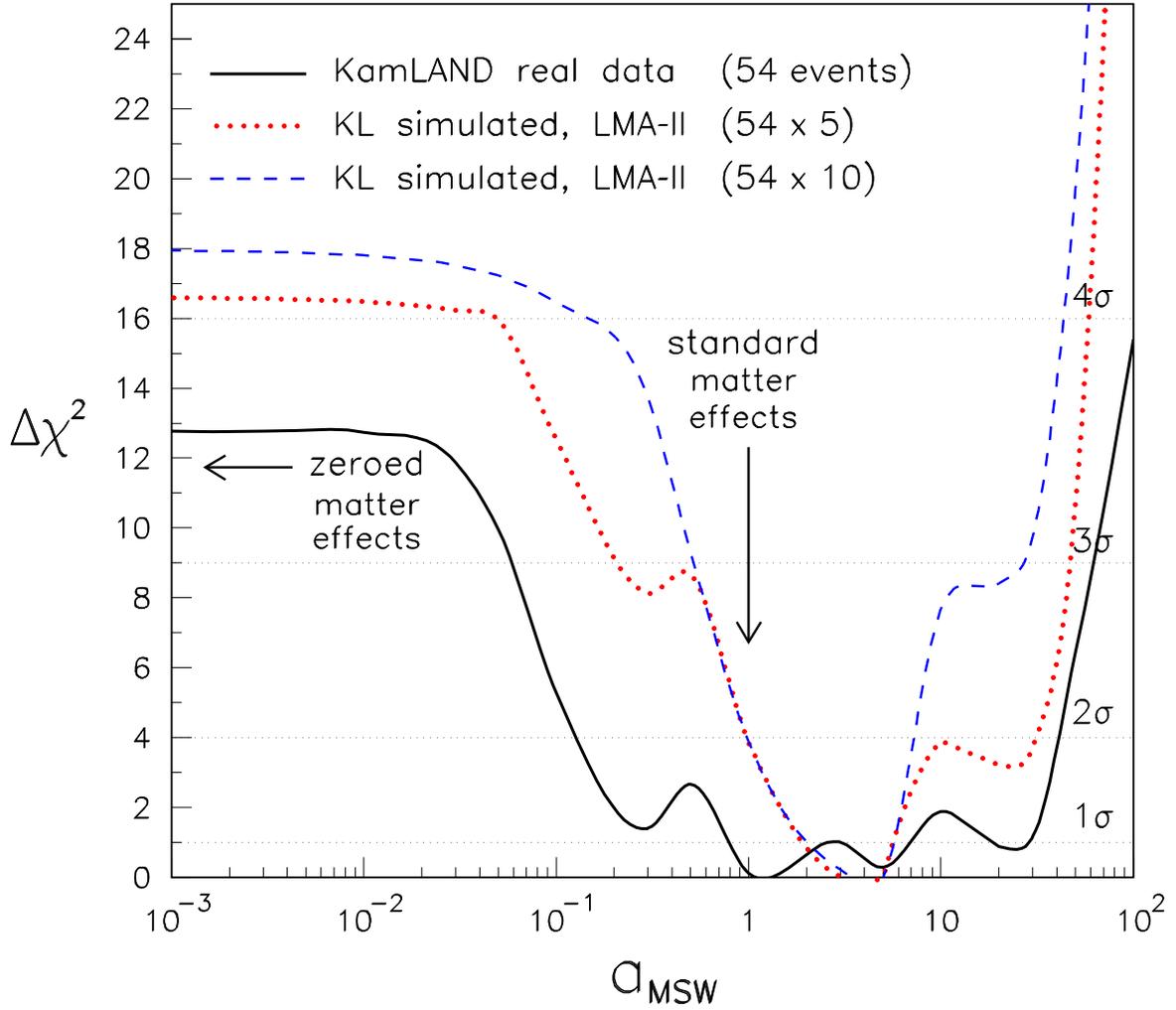}
\vspace*{-1cm} \caption{\label{fig3} As in Fig.~2, except that the
dotted and dashed curves here refer to simulated KamLAND data for
the LMA-II solution of Ref.~\cite{KLou}. This solution would imply
a mismatch between the reconstructed best-fit values of $\delta
m^2$ obtained separately from current solar data and prospective
KamLAND data. The mismatch is reflected here through a
reconstructed amplitude of matter effects $a_\mathrm{MSW}$
typically greater than (although still compatible with) standard
expectations.
 See the text for details.}
\end{figure}

\end{document}